\documentclass[pra,twocolumn,amsmath]{revtex4-1} \usepackage{amssymb} \newcommand{\tr}{{\rm Tr}} 

\pdfoutput=1

\newcommand{\relevant}{relevant}
\newcommand{\irrelevant}{irrelevant}
\newcommand{\relevance}{relevance}

\usepackage{amsthm}
\usepackage{graphicx}
\usepackage[all]{xy}
\usepackage[usenames,dvipsnames]{color}
\usepackage{tabularx}

\newcommand{\mc}[1]{\mathcal #1}

\newcommand{\ave}[1]{\langle #1 \rangle}
\newcommand{\bra}[1]{\langle #1 |}
\newcommand{\ket}[1]{| #1 \rangle}

\newcommand{\one}{{\bf 1}}

\renewcommand{\eqref}[1]{(\ref{#1})}

\newcommand{\bigave}[1]{\bigl \langle #1 \bigr \rangle}

\newcommand{\prlsection}[1]{\section{#1}}

\newcommand{\half}{\frac{1}{2}}

\theoremstyle{plain}

\theoremstyle{definition}

\begin{document}

\title{The renormalisation group via statistical inference}
\author{C\'edric B\'eny and Tobias J.\ Osborne}
\address{Institut f\"ur Theoretische Physik, Leibniz Universit\"at Hannover, Appelstra{\ss}e 2, 30167 Hannover, Germany}
\date{\today}

\begin{abstract}
In physics one attempts to infer the rules governing a system given only the results of imperfect measurements. Hence, microscopic theories may be effectively indistinguishable experimentally. We develop an operationally motivated procedure to identify the corresponding equivalence classes of states, and argue that the renormalisation group arises from the inherent ambiguities associated with the classes: one encounters flow parameters as, e.g., a regulator, a scale, or a measure of precision, which specify representatives in a given equivalence class. This provides a unifying framework and reveals the role played by information in renormalisation. We validate this idea by showing that it justifies the use of low-momenta $n$-point functions as statistically relevant observables around a gaussian hypothesis. These results enable the calculation of distinguishability in quantum field theory. Our methods also provide a way to extend renormalisation techniques to effective models which are not based on the usual quantum-field formalism, and elucidates the relationships between various type of RG. 
\end{abstract}


\maketitle

The renormalisation group (RG), as conceived by Wilson \cite{wilson:1974a,wilson:1975a}, relies on the idea that it is possible to describe long-distance physics while essentially ignoring short-distance phenomena; Wilson argued that, if we are content with predictions to some specified accuracy, the effects of physics at smaller lengthscales can be absorbed into the values of a few parameters of some \emph{effective theory} for the long-distance degrees of freedom. 
The RG now underpins much of our understanding of modern theoretical physics and provides the interpretational framework for quantum field theories. It has been applied in a dazzling array of incarnations to study systems from statistical physics \cite{fisher:1998a} to applied mathematics \cite{barenblatt:1996a}. 

The general applicability of RG techniques strongly suggests the existence of a deep unifying principle which would make it possible to directly compare different manifestations of the RG and to unlock its full potential.
It has been suggested that such a general implementation-independent formulation of the RG is to be found in an information-theoretic approach~\cite{preskill:2000a} because the RG works by {\em ignoring} certain aspects of the system. Although the information-theoretic flavour of the RG is manifest in the case of block-decimation~\cite{kadanoff:1976a, kadanoff:1966a, kadanoff:1977a}, it is far less obvious in the context of particle physics from where the terminology of renormalisation originates~\cite{zamolodchikov:1986a}. 
Previous attempts at tackling this problem (see, e.g., \cite{apenko:2012a,brody:1998a,casini:2007a,gaite:1996a,heckman13} for a selection) depend on details of the chosen model or formalism and do not yet offer the truly general unification that one might hope for. 

The objective of this paper is to propose an operationally motivated, model-independent, and hence information-theoretic framework for the RG. Our main result is a demonstration that this framework encompasses, as a particular case, the renormalisation group implemented with respect to a regulator (as found in QFT).

Our approach is related to that of a recent paper of Machta et al.~\cite{machta13}, who observed that the relevant parameters selected by the RG have the property that they generate perturbations of a statistical state which are distinguishable (in information-theoretic terms) even when the system is coarse-grained. This is why these parameters can be {\em inferred} experimentally and are useful for predictions. 

We first step back, and phrase the inference task as a game played between two players: a passive one, Alice, who simply possesses a quantum or classical system, and Bob, who perceives the system via a known noisy quantum channel $\mc E$. (That is, any map linearly taking density matrices to density matrices, even as part of a larger system. Classically, it is any stochastic map). The channel may for instance represent a coarse-graining.
We think of Alice as possessing the true state of a physical system, while Bob is an experimentalist whose practical limitations are formalised by the channel.  When Bob tries to infer the state of Alice's system, he is faced with the ill-posed \emph{inverse problem} of inverting a quantum channel to find the input from the output. 

\begin{figure}
\begin{center}
\begin{tabularx}{0.8\columnwidth}{XllX} (a) & (b) \\
\includegraphics[width=0.25\columnwidth]{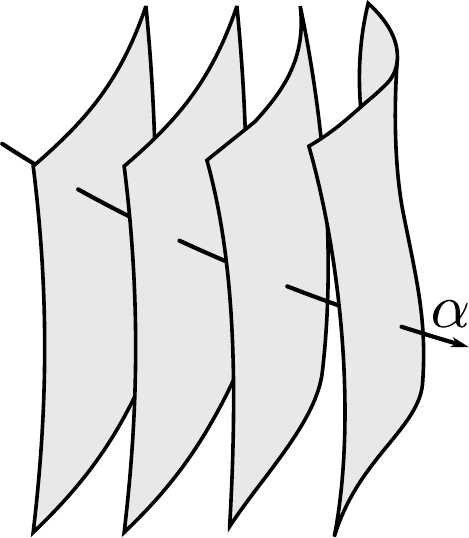} & \includegraphics[width=0.25\columnwidth]{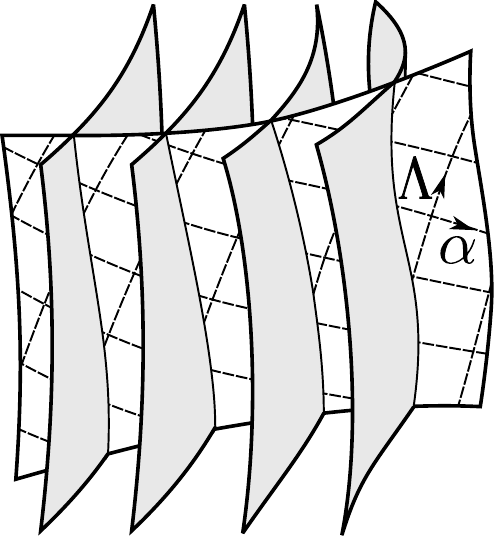} \\
\end{tabularx} \caption{
The shaded planes represent equivalence classes of states which cannot be distinguished experimentally. They are intersected by the manifold of effective theories, parametrised in example (a) by a sole parameter $\alpha$, and, in example (b), additionally by a regularisation parameter $\Lambda$. The intersection lines are renormalisation trajectories $\alpha(\Lambda)$. 
}
\label{foliation}
\end{center}
\end{figure} 

Let us consider first a situation where the channel has a non-trivial kernel. For instance, $\mc E$ could be the partial trace over all high-momentum modes of a theory. If two states $\rho$ and $\rho'$ are such that $\mc E(\rho)=\mc E(\rho')$ then they cannot be distinguished by Bob and hence are both just as good as hypotheses for Alice's state. This indistinguishability results in \emph{equivalence classes} of states: all that Bob can hope to do is to determine in which class the true state is. The classes can be parameterised by a smooth manifold of unique representatives (Fig.~\ref{foliation}a).
For instance, if $\mc E$ traces out high-momentum modes, the equivalence classes can be labelled by states whose high momentum modes are in some fiducial product state. 

Once the classes of experimentally indistinguishable states are identified, we propose that the various existing types of RG result from an exploration of the freedom available in choosing the representative within a class. For example, when modifying a regularisation parameter, as occurs in high-energy physics, or when simplifying the description of the state and isolating the relevant degrees of freedom, as commonly practised in condensed matter theory. Before we describe these two cases in more depth, we need to consider more general, and more realistic experimental limitations. This requires taking {\em approximate} indistinguishability into account.

\prlsection{General framework}
A reasonable measure of distinguishability between two states $\rho$ and $\rho'$ to be used in this situation is the relative entropy 
\begin{equation}
S(\rho' \|\rho) = {\rm Tr}\left({ \rho\, ( \log \rho -  \log \rho') }\right),
\end{equation} 
which measures the optimal exponential rate of decrease of the probability of mistaking $\rho'$ for $\rho$ as a function of the number of copies available, while still letting the probability of mistaking $\rho$ for $\rho'$ go to zero~\cite{hiai:1991}. This asymmetric scenario is relevant to the situation where one attempts to prove the new hypothesis $\rho'$ against a well established one: $\rho$. 
Our framework can also be adapted to different measures, but we use this one here for concreteness. 
The above interpretation is for an observer able to measure any observable on Alice's system. Bob, however, has a limited access to Alice's state. Since he can only make direct measurements on the states $\mc E(\rho)$ and $\mc E(\rho')$, his optimal ability to distinguish between $\rho'$ and $\rho$ according to the above scenario is instead given by the rate $S(\mc E(\rho')\|\mc E(\rho))$.

The effect of $\mc E$ can also be thought of as limiting the type of observable that Bob can measure directly on Alice's system, through the {\em Heisenberg picture} defined via the adjointness relation $\tr(\rho \, \mathcal{E}^\dag(A)) = \tr(\mc E(\rho) A)$: Bob can effectively only measure POVMs on $\rho$ with elements $\mathcal{E}^\dag(A_i)$, where $0 \le A_i \le \one$ and $\sum_i A_i = \one$. His effective distinguishability rate $S(\mc E(\rho')\|\mc E(\rho))$ is hence smaller than that of an all powerful experimentalist, namely $S(\rho'\|\rho)$, because he has access to fewer observables.


Consequently, we could attempt to deem two states $\rho$ and $\rho'$  experimentally equivalent if $S(\mc E(\rho')\| \mc E(\rho))<\delta$ for some desired maximal rate $\delta$. However, this does not define an equivalence relation (this relation is not transitive, nor even symmetric). 
Nevertheless, if $\delta$ is sufficiently small, we still expect that the set of states $\rho'$ close to $\rho$ form an approximately linear subspace of matrices, as occurs in the exact case $\delta=0$. This motivates us to \emph{linearise} the relation around a starting hypothesis $\rho$. 

Let us consider the state $\rho' =\rho+\epsilon X$, where $\epsilon$ may be arbitrarily small. We will call the operator $X$, which must be hermitian and traceless, a {\em feature}. In terms of the manifold of density matrices, $X$ represents a tangent vector to the point $\rho$. (It is related to the tangent vector represented as differential operator $\hat X$ on scalar functions $f$ via $f(\rho + \epsilon X) = f(\rho) + \epsilon (\hat X f)(\rho) + \mc O(\epsilon^2)$). 

Then, to lowest order in $\epsilon$, we have 
\begin{equation}
S(\rho+\epsilon X \| \rho)\; \approx \;\epsilon^2 \tr(X \,\Omega_\rho^{-1}(X)) \;\equiv\; \epsilon^2 \ave{X,X}_\rho, 
\end{equation}
where $\Omega_\rho^{-1}(Y)  = \frac{d}{dt}\log(\rho+ t Y) |_{t=0}$ is a non-commutative version of the operation ``division by $\rho$''. 
The quantity $\ave{X,Y}_\rho\equiv\tr(X \,\Omega_\rho^{-1}(Y)) $ is an inner product on operators. Since it is defined at every point of the manifold of states, it is a \emph{metric} in the sense of differential geometry and is one of the many quantum generalisations of the Fisher information metric \cite{petz:1996a}.

In this linear approximation, a state $\rho + X$ is approximately indistinguishable from $\rho+Y$ by Bob if 
\begin{equation}
\ave{\mc E(X-Y),\mc E(X-Y)}_{\mc E(\rho)} < \delta.
\end{equation}
The set of states $\rho+X$ satisfying this condition is an ellipsoid within Alice's state space. If $\mc E$ is not invertible, the ellipsoid is infinitely wide in the null directions $Z$ with $\mc E(Z) = 0$.  Consequently, in the generic case, we use the following idealised relation: we say that the two states $\rho +X $ and $\rho + Y$ are equivalent if $X-Y$ lies in the span of the ``largest'' principal directions of the ellipsoid (those that contract ``the most'').

This idealisation removes any trace of the desired precision $\delta$, as we are only talking of the direction of $X-Y$ independently of its magnitude. Instead, Bob must choose the number $n$ of features he deems sufficiently distinguishable. A pertinent way of doing this is to consider the case where the channel $\mc E$ depends on a parameter $\sigma$ parameterising the precision of Bob's instruments, and to worry about the asymptotic behaviour of the norm  $\ave{\mc E(Z),\mc E(Z)}_{\mc E(\rho)}$ in the limit of large imprecision $\sigma$. The choice of threshold $n$ then amounts to choosing the type of asymptotic behaviour that we deem negligible. This is what happens in the examples presented below.

The principal directions of the ellipsoid are obtained by a singular value decomposition of $\mc E$ with respect to the scalar product defined by the metric.  
Let $\mc R_\rho$ be the {\em adjoint} of $\mc E$ defined by $\ave{X,\mc R_\rho(Y)}_\rho = \ave{\mc E(X),Y}_{\mc E(\rho)}$ for all features $X, Y$. Explicitly, 
\begin{equation}
\mc R_\rho = \Omega_\rho \mc E^\dagger \Omega_{\mc E(\rho)}^{-1}.
\end{equation}
This map generalises the {\em transpose channel}~\cite{ohya04}. Classically, if $p(y|x)$ are the components of $\mc E$, then $\mc R_\rho$ has for components the conditional probabilities $p(x|y)$ derived from Bayes' rule with prior $\rho$.
The {\em principal features} $X_j$ are the solution of 
\begin{equation}
\mc R_\rho\mc E(X_j)=\eta_j X_j.
\end{equation}
The eigenvalues $\eta_j$ are also the singular values of $\mc E$, and satisfy $1\ge\eta_1\ge\eta_2\ge\dots\ge 0$. We call $\eta_j$ the {\em \relevance{}} of $X_j$.
The linear operator $\mc R_\rho \mc E$ is self adjoint in the scalar product $\ave{\cdot,\cdot}_\rho$. Therefore, the principal features form an orthogonal basis of the tangent space at $\rho$. 

This concept of \relevance{} is a genuinely coordinate independent version of the {\em stiffness} defined in Ref.~\cite{machta13}. It equals stiffness computed with respect to the special parametrisation in which the original metric is given by the identity matrix. 

We call a feature $X$ {\em \relevant{}} if it is in the span of $X_1,\dots, X_n$, and {\em \irrelevant{}} if it is orthogonal to those.
Our idealised equivalence classes make $\rho + X$ and $\rho +Y$ equivalent from the point of view of Bob if and only if $X-Y$ is \irrelevant{}, or, equivalently, if $\ave{X-Y,Z}_\rho = 0$ for all \relevant{} feature $Z$. 

In order to obtain a physically more intuitive condition, let us define the {\em principal observables} to be the operators $A_j = \Omega_\rho^{-1}(X_j)$, solutions of the dual Heisenberg-picture eigenvalue equation 
\begin{equation}
\label{eq:eigen}
\mc E^\dagger \mc R_\rho^\dagger(A_j)  = \eta_j A_j.
\end{equation}
Analogously, we say that $A$ is a \relevant{} observable if it is in the span of $A_1, \dots, A_n$.
With this definition, our equivalence condition amounts to considering two effective states $\rho'$ and $\rho''$ to be equivalent (in the neighbourhood of $\rho$) when they yield the same expectation values {\em for all \relevant{} observables}:
\begin{equation}
\rho' \sim_\rho \rho'' \quad \text{iff} \quad \tr(\rho' A_j) = \tr(\rho'' A_j) \quad \text{for} \quad j \le n.
\end{equation}

For instance, consider the strictest possible \relevance{} threshold where only features with exactly zero \relevance{} are deemed to be \irrelevant{}. These are the operators $X$ in the kernel of $\mc E$. In this case we recover the exact state-independent equivalence relation which identifies $\rho'\sim \rho''$ if $\mc E(\rho')=\mc E(\rho'')$. The corresponding \relevant{} observables are the self-adjoint operators $A$ satisfying $\tr(AX) = 0$ for all $X$ in the kernel of $\mc E$, which are precisely those of the form $\mc E^\dagger(B)$ for some $B$. 
In addition, these are all the observables that Bob can ever hope to measure expectation values of, since for all $B$, $\tr(B\mc E(\rho))=\tr(\mc E^\dagger (B)\rho)$. 

\prlsection{One classical mode}

For a simple but nontrivial example suppose that Alice has a stochastic classical system consisting of a single real variable, e.g., the position $x$ of a particle. The true state to be discovered by Bob is a probability distribution $\rho(x)$. Bob's experimental limitation consists of a finite precision $\sigma$ at which he can resolve $x$. This can be modelled by a stochastic map $\mc E$ whose effect is a convolution of Alice's probability distribution with a Gaussian of width $\sigma$:
\begin{equation}
\mc E(\rho)(x) = \int \rho(y) N_\sigma(x-y) dy
\end{equation}
here $N_\sigma$ is the normal distribution with variance $\sigma$. 
Suppose, further, that Bob's initial hypothesis is a simple gaussian distribution, which we think of as a thermal state $\rho(x) \propto e^{-H(x)}$ for the ``hamiltonian'' $H(x) = \frac{x^2}{2\tau^2}$. The action of the operator $\mc R_\rho^\dagger$ can be written explicitly:
\begin{equation}
\mc R_\rho^\dagger(f)(x) = \frac{ \int dy f(x)  \rho(x) N_\sigma(x-y) dy }{\int dy \rho(x) N_\sigma(x-y) dy}.
\end{equation}
Noting also that $\mc E = \mc E^\dagger$, one can check by explicit calculation of the gaussian integrals that if $G_t(x) = e^{t x /\tau - t^2/2}$, then
\begin{equation}
\label{eq:onepartsol}
\mc E^\dagger \mc R_\rho^\dagger(G_t) = G_{\eta t}
\end{equation}
with
\(
\eta = \frac{\tau^2}{\sigma^2 + \tau^2}.
\)
Hence, the eigenvalue problem defined in Equ.~\eqref{eq:eigen} is solved by differentiating Equ.~\eqref{eq:onepartsol} $n$ times with respect to $t$, evaluated at $t = 0$. Observe that $G_t(x)$ is the generating function for the Hermite polynomials, hence the principal observables are the hermite polynomials ${\rm He}_n(x/\tau)$, with respective eigenvalues $\eta_n =\eta^n$, or $\eta_n\approx(\tau/\sigma)^{2n}$ for $\sigma\gg\tau$. 
Following our criterion this means that, since the first $n$ hermite polynomials span all polynomials of degree $n$, that for a threshold $n$, two nearby states are equivalent exactly when they share the same first $n$ moments.

For instance, up to distinguishability of order $\sigma^{-4}$ $(n=2)$, the effective hamiltonian $H_0(x) = \frac{x^2}{2\tau_0^2} + \lambda x^4$ is equivalent to $H_1(x) = \frac{x^2}{2\tau_1^2}$ provided that $\tau_1$ is ``renormalised'' so as to yield the same second moment as $H_0$. This simplification from $H_0$ top $H_1$ morally corresponds to a step of the type of RG employed in condensed matter theory, where a hamiltonian is simplified in a way that only affects some ``unobservable'' small scale features of the systems. 

The situation in particle physics is a priori quite different. Quantum field theories typically come with an unwanted parameter, a \emph{regulator} $\Lambda$, which has no true physical significance, although it often mimics a lattice spacing. Its presence, however, is not a problem if the observable predictions of the theory do not depend on it. This is possible if we assume some reasonable limitation on Bob's measurement abilities, so that any change in $\Lambda$ can be compensated by a change in the state's other parameters so as to stay within a given equivalence class (Fig.~\ref{foliation}b).
This dependance of the state's parameters on $\Lambda$ is the type of RG flow which naturally occurs in QFT. 

Using the above toy example, a similar problem could occur for the hamiltonian $H_0$ if $\lambda$ were to be experimentally determined to be negative (using a first order approximation in $\lambda$ for the state). Indeed, the resulting distribution $\rho(x)$ would diverge if calculated non-perturbatively. This can be fixed mathematically by adding a sixth order term $x^6/\Lambda$ to the effective hamiltonian, which, to distinguihsability of order $\sigma^{-4}$, can be made to be equivalent to $H_0$ by adjusting the parameters $\tau$ and $\lambda$ as function of $\Lambda$ so as to preserve up to the fourth moment. 

Those two concepts of renormalisation can be made to match in QFT because divergences can be identified as contributions from an infinite number of {\em \irrelevant{}} features. Hence, the simplification which consists in subtracting them from the state also regularises the theory.

\prlsection{Classical gaussian states}

We solve Equ.~\eqref{eq:eigen} for gaussian states over arbitrarily many modes, and for a channel $\mc E$ which is any gaussian stochastic map. We consider $n$ real random variables $\phi_i$. We write $\phi(f) := \sum_i f_i \phi_i$ for any vector $f$, which corresponds to a ``smeared'' field in the continuum limit.
A general gaussian stochastic map $\mc E$ is defined by the effect of its transpose to the moment-generating functions:

\begin{equation}
\mc E^\dagger(e^{\phi(f)})=e^{\phi(X f)+\half(f,Y f)},
\end{equation}
where $(f,g)=\sum_if_ig_i$, $Y$ and $X$ are real matrices and $Y$ is positive (we give a concrete example in the quantum case).
Similarly, an arbitrary (but centered) gaussian state $\rho$ is defined by
\begin{equation}
\ave{e^{\phi(f)}}_\rho=e^{\half(f,A f)}
\end{equation}
where $A$ is real and symmetric.
Using the definition of $\mc R_\rho^\dagger$ as adjoint of $\mc E^\dagger$ in the dual metric, applied to the generating functions $e^{\phi(f)}$, one can show that the random variables
\begin{equation}
G(f):=e^{-\half(f,Af)+\phi(f)}
\end{equation}
satisfy
\begin{equation}
\mc E^\dagger\mc R^\dagger_{\rho}(G(f))=G(Hf).
\end{equation}
with 
\begin{equation}
H = (1 + A^{-1} X^{-1} Y X^{-1})^{-1}
\end{equation} 
(See Appendix A). Note that $H$ is symmetric with respect to the scalar product $(\,\cdot\,, A \,\cdot\,)$, hence it has a complete orthonormal family of eigenfunctions $f_k$ with eigenvalues $\eta_k$. We obtain the eigen-variables of $\mc E^\dagger \mc R^\dagger_{\rho}$ (namely the principal observables) explicitly by differentiating the generating functional $G$ in the directions of the functions $f_k$ any number of times, and evaluating the result at $f = 0$.

\prlsection{Interactions}
The previous result can be used to perturbatively calculate the principal observables around nongaussian states. In order to do this, we need to work within a representation of the real Hilbert space formed by the principal observables of the gaussian state, together with the scalar product defined by the metric evaluated at the gaussian state. 
This is always a symmetric Fock space, where the vacuum $\ket 0$ corresponds to the constant random variable $G(0) = 1$ (with \relevance{} $1$), and
the creation operator $a_k^\dagger$ associated with vector $f_k$, acting on a principal observable, leads to a new principal observable with \relevance{} multiplied by $\eta_k$. 
The perturbed eigenvalue problem can then be expressed  
to any order using standard Feynman diagrams.
An example is detailed in the Appendix B. We show below, however, that the standard RG conditions in QFT can be recovered from the gaussian results alone. 

\prlsection{Quantum gaussian states}
A quantum gaussian state is defined by quantization of a classical phase space. Let $f$, $g$ denote classical observables which are linear functions in the canonical variables, with some scalar product $(f,g)$ and the symplectic form $\Delta$. Let $\Phi(f)$ denote the quantisation of $f$, such that
\begin{equation}
[\Phi(f),\Phi(g)]=i(f,\Delta g)\one.
\end{equation}
Any quantum state is uniquely specified by its characteristic function $f \mapsto \ave{e^{i\Phi(f)}}_\rho$. For a quantum gaussian state $\rho$, this is $e^{-\half (f, Af)}$, where $A$ is a real symmetric matrix satisfying $A + \frac i 2 \Delta \ge 0$. 
A general gaussian channel is characterised by its effect on the Weyl operators:
\begin{equation}
\label{eonweyl}
\mc E^\dagger(e^{i\Phi(f)})=e^{i\Phi(Xf)}e^{-\half(f,Yf)},
\end{equation} 
where $X$ and $Y$ are real matrices such that 
\begin{equation}
\label{eq:chancond}
Y-\frac{i}{2} X^\dagger\Delta X+\frac{i}{2} \Delta\ge 0.
\end{equation}

One can then verify that the principal observables are polynomials generated by 
\begin{equation}
G_A(f)=e^{\frac{1}{2}(f, Af)+i\Phi(f)}.
\end{equation}
This is done by first noting their orthogonality, and applying the definition of $\mc R_\rho$ with respect to the generating function as in the classical case (See Appendix C). 

As an example, we consider a gibbs state of a Klein-Gordon field of mass $m$ at inverse temperature $\beta$, with canonical conjugate coordinates $\phi(x)$ and $\pi(x)$. We will need the real fourier components
\begin{eqnarray}
\phi_k &=\int dx[\cos(kx)\phi(x)-\omega_k^{-1}\sin(kx)\pi(x)]\\
\pi_k &=\int dx[\omega_k\sin(kx)\phi(x) + \cos(kx)\pi(x)],
\end{eqnarray}
which are decoupled under the classical dynamics. 
Because the phase space is infinite-dimensional, the concept of gaussian state introduced above has to be generalised with some care. Alternatively, one may choose boundary conditions and a momentum cutoff so as to render it finite-dimensional. 
For our purpose, we define the gaussian state through the bilinear form that it defines on the space of linear classical observables. In terms of the observables
\begin{eqnarray}
f(\phi,\pi) &= \int dk \, ( f_k \phi_k  + f_k' \pi_k )\\
g(\phi,\pi) &= \int dk \, ( g_k \phi_k  + g_k' \pi_k ),
\end{eqnarray}
the quadratic form is
\begin{equation}
(f, A g) = \frac 1 2 \int dk \, \coth\bigl( \beta \omega_k/2\bigr)\Bigl({ \omega_k f_k g_k + \frac 1 {\omega_k} f'_k g'_k }\Bigr).
\end{equation}
We also consider a gaussian channel $\mc E$. In the continuum, the matrices $X$ and $Y$ become linear functions. We use 
\begin{equation}
\label{Xdef}
(Xf)(\phi,\pi) = f(N_\sigma\star\phi,N_\sigma\star\pi),
\end{equation}
where $N_\sigma\star\cdot$ denotes convolution by a gaussian of variance $\sigma$, and 
\begin{equation}
(Yf)(\phi,\pi)=f(y_\phi\phi,y_\pi\pi).
\end{equation}
The parameter $\sigma$ characterises spatial resolution, and $y_\phi$ and $y_\pi$ field value resolutions.
The condition expressed in Equ.~\ref{eq:chancond} reduces in this case to the uncertainty relation $y_\phi y_\pi \ge 1$.

For $y_\phi y_\pi\gg 1$, we find that the quantized field observables $\hat\phi_k = \Phi(\phi_k)$ and $\hat\pi_k = \Phi(\pi_k)$ are principal observables, with respective \relevance{}
\begin{equation}
\eta^\phi_k \simeq \frac 1 {\frac{\beta\omega_k}2 \coth \frac{\beta\omega_k} 2 +\beta\omega_k^2 y_\phi^{2}\,e^{k^2\sigma^2}}
\end{equation}
and
\begin{equation}
\eta^\pi_k  \simeq \frac 1 {\frac{\beta\omega_k}2 \coth \frac{\beta\omega_k} 2 +\beta y_\pi^{2}\,e^{k^2\sigma^2}}.
\end{equation}
The solution for higher order polynomials in the field operators is more complex. However, one can show that the relevance of any polynomial of order $n$ in the modes $k_1, \dots, k_n$ is bounded by a constant times $e^{- \sum_{i=1}^n k_i^2 \sigma^2}$. Moreover, assymptotically in terms of $y^2 := y_\phi y_\pi$, with $y_\phi/y_\pi$ constant, the relevance of these polynomials also scale like $y^{-2n}$. This is shown in Appendix C.

\prlsection{Renormalisation}
If we want to recover a renormalisation group, we have to pick a threshold on the asymptotic decay of \relevance{} in terms of the three noise parameters $\sigma$, $y_\pi$ and $y_\phi$. Since the \relevance{} decays exponentially in the total momentum, the product of field operators can always be considered \irrelevant{} as soon as they involve operators with mode $k\gg 1/\sigma$.  
At temperatures large compared to $m$, the \relevance{} of the $n$-point functions also decays to order $2n$ in $y_\pi$ and $y_\phi$. Hence, in this approximation two states are effectively equivalent if they have the same $n$-point functions at momenta smaller than $1/\sigma$. Without restriction on $n$, this is precisely the conditions used in QFT for the RG as function of a regulator $\Lambda$ (needed as momentum cutoff on divergent integrals resulting from perturbation theory around gaussians). 
Instead of viewing the cutoff as an explicit parameter of the state, a change of cutoff from $\Lambda$ to $\Lambda'$ can also be absorbed into a {\em rescaling} of space by a factor $s=\Lambda'/\Lambda$. The condition that the state stays in the same equivalence class independantly of $s$ yields the Callan-Symanzik equations.

From the condensed matter point of view, a cutoff $\Lambda$ is fixed and given by the lattice spacing. The description of the state can be simplified by exploiting the freedom we have in choosing a representative of an equivalence class. We may chose the one closest to $\rho$ in relative entropy: this optimisation is well known and yields the gibbs states with only \relevant{} observables in the hamiltonian perturbations, namely terms with field modes $|k| < 1/\sigma$. The requirement that visible predictions be invariant from $\sigma$ leads to an RG. This matches the previous RG in the sense that the procedure is technically equivalent to lowering the cutoff $\Lambda$ to $1/\sigma$ in perturbative expansions.

But our calculation also tells us that we may neglect features whose distinguishability scales poorly with the field-value precisions $y_\phi$ and $y_\pi$, hence justifying the use of effective hamiltonians with low degree polynomials in the fields. For instance, choosing $n=2$ selects only quadratic terms as \relevant{} observables, and the gaussians states as a natural family of effective states (this is a very different argument than the one based on renormalisibility). We note that here $y_\phi$, $y_\pi$ play very different roles than $\sigma$ because of the differences in asymptotic \relevance{} behaviour, but we could imagine a different type of experimental limitations where more parameters govern the RG besides $\sigma$.

\prlsection{Distinguishability in QFT}
We can use the solution of Equ.~\eqref{eq:eigen} to compute the effective \emph{distinguishability} $D(A)$ of any perturbation generated by a Hamiltonian term $A$, defined as the lowest-order approximation of the relative entropy $S(\mc E(\rho_\epsilon)\|\mc E(\rho_0))=\epsilon^2 D(A)+\mc O(\epsilon^3)$, where $\rho_\epsilon \propto e^{-H+\epsilon A}$ are normalised states. Indeed, we have $D(A)=\ave{A,\mc E^\dagger\mc R_\rho^\dagger(A)}_\rho$, which can be computed by expressing $A$ in the basis of principal observables around $\rho$. For instance, in the scalar field example, $D( \hat \phi_k)=\eta_k^\phi \ave{\hat \phi_k, \hat \phi_k}_{\rho}$, where $\eta_k^\phi$ is given above, and $\ave{ \hat \phi_k, \hat \phi_k}_{\rho}=1/\beta\omega_k^{2}$. With the standard tools of perturbative QFT, this can be generalised to higher-order expansions of the exponential (while keeping with the lowest order approximation of the relative entropy).

For non-local terms, $D$ has to be made into a density. It may be argued that the unit of volume used to define the density ought to explicitely {\em scale} with $\sigma$, leading to the distinguishability density $d_\sigma(A)=\sigma^d\lim_{\Sigma}D_\sigma(A_\Sigma)/|\Sigma|$, where $d$ is the dimension of space and $A_\Sigma$ a restriction of $A$ to a region $\Sigma$ of volume $|\Sigma|$. A Hamiltonian term may then be said to be {\em relevant} in information-theoretic terms if $d_\sigma$ scales as a positive power of $\sigma$ (for a fixed state). Preliminary calculations indicate that the result is compatible with the Wilsonian analysis classically, but may differ in important ways in the genuinely quantum analysis. This will be analysed in further work.

\prlsection{Concluding remarks}
We introduced a framework which allows for the definition of effective theories in very general terms, taking into account any measure of distinguishability and any model of experimental limitations. We demonstrated the pertinence of this approach by showing that it naturally contains, as a particular case, the concept of effective theory as defined by the renormalisation group of quantum field theory. Further work will explore how varying the assumptions lead to effective theories which differ from the standard QFT framework. For instance, taking the field-value resolutions into account in the interacting context leads to a concept of dressed effective field which depends in principle on the detail of the coarse-graining channel and distinguishability metric. 

Most interestingly, the fact that this approach is not at all tied to the standard QFT formalism means that it can in principle be applied to completely different types of theories, as well as very different models of experimental limitation (not necessarily related to scale). For instance, the case of loop quantum gravity~\cite{rovelli14}, which proposes a class of background-free quantum field theories, could provide interesting applications. 

This approach can also be naturally applied to spin lattice systems, so as to derive effective field theories describing their large scale properties. In standard approaches, a spins system is connected to an effective QFT through symmetry arguments (observables in the discrete and continuous descriptions are paired by identifying the group transformations they generate). Our approach provides a more bottom-up approach, where the effective QFT can in principle be derived through the mechanism by which it emerges, i.e., through the identification of the degrees of freedom which are effectively ignored. This can be performed numerically by solving Equ.~\ref{eq:eigen} using techniques such as matrix product states. 

Finally, the framing of effective QFT in this fundamentally information-theoretical approach elucidates precisely what information is being destroyed when the theory is renormalised. A concrete way of quantifying this is proposed in the last section. This may provide a first step towards generalising Zamolodchikov's c-theorem~\cite{zamolodchikov:1986a,beny12}, which could in turn provide new techniques for the general classification of effective QFTs. 

\section*{Acknowledgments}
Helpful discussions with numerous people are most gratefully acknowledged: a partial list includes John DeBrota, Andrew Doherty, Jens Eisert, Steve Flammia, Jutho Haegeman, Gerard Milburn, Terry Rudolph, Tom Stace, Frank Verstraete, and Reinhard Werner. This work was supported by the ERC grants QFTCMPS and SIQS and by the cluster of excellence EXC 201 Quantum Engineering and Space-Time Research.

\section*{Appendix A: Classical gaussian states}

We want to find the action of $\mc R^\dagger_{\rho'}$ on $G(f)$. By directly using its definition as adjoint of $\mc E^\dagger$, we find that for any linear classical observables $f$ and $g$,
\begin{eqnarray*}
&\ave{\mc R_{{\rho'}}^\dagger(e^{\phi(f)}),e^{\phi(g)}}_{\mc E({\rho'})} = \ave{e^{\phi(f)},\mc E^\dagger(e^{\phi(g)})}_{\rho'}\\
&\quad \quad = e^{\frac 1 2 (g, Y g)} \ave{e^{\phi(f + X g)}}_{\rho'}\\
&\quad \quad = e^{\frac 1 2 (g, Y g) )+ \frac 1 2 (f + X g,A (f + Xg))}\\
&\quad \quad = e^{\frac 1 2 (f,A f)  + \frac 1 2 (g,(X A X + Y) g) + (g, XA f)}\\
\end{eqnarray*}
But note that, using
\begin{equation}
j := (XAX + Y)^{-1} X A f,
\end{equation}
we have
\begin{eqnarray*}
\ave{e^{\phi(j)},e^{\phi(g)}}_{\mc E(\rho)} &= e^{ \frac 1 2 (j,Y j)  + \frac 1 2 (g,(XAX + Y)g) + (g,X A f)} \\
& \quad \times e^{\frac 1 2 (j, XAX j)}.
\end{eqnarray*}
By comparing the two expressions, we obtain
\begin{eqnarray*}
\langle\mc R_{\rho}^\dagger(e^{\phi(f)}), e^{\phi(g)}\rangle_{\mc E(\rho)}
&= e^{\frac 1 2 (f,A f) -\frac 1  2 (j, Yj) - \frac 1 2 (j, XAX j)} \\
&\quad \times \ave{e^{\phi(j)},e^{\phi(g)}}_{\mc E(\rho)}.   \\
\end{eqnarray*}
Since this is true for all $g$, then,
\begin{eqnarray*}
\mc R_{\rho}^\dagger(e^{\phi(f)}) &= e^{\frac 1 2  (f,A f)  - \frac 1 2 (j, (XAX + Y) j)} e^{\phi(j)}\\
\end{eqnarray*}
or
\begin{equation}
e^{-\frac 1 2  (f,A f)} \mc R_{\rho}^\dagger(e^{\phi(f)}) = e^{ - \frac 1 2 (j, (XAX + Y) j)} e^{\phi(j)}\\
\end{equation}
and hence
\begin{equation}
\mc E^\dagger \mc R_{\rho}^\dagger(e^{-\frac 1 2  (f,A f) + \phi(f)}) = e^{ - \frac 1 2 (X j, A X j)} e^{\phi(X j)}.
\end{equation}
Note that
\begin{equation}
Xj =  (1 + A^{-1} X^{-1} Y X^{-1})^{-1}  f.
\end{equation}
Hence, defining $H = (1 + A^{-1} X^{-1} Y X^{-1} )^{-1}$, 
we conclude that
\begin{equation}
\mc E^\dagger \mc R_{\rho}^\dagger(G(f)) = G(Hf).
\end{equation}

Observe that $A H = H^T A$. Hence $H$ is symmetric in the scalar product $(\cdot, A \cdot)$.
Let $f_k$ be am orthonormal basis of eigenfunction of $H$:
\begin{equation}
H(f_k) = \eta_k f_k \quad\text{and} \quad (f_k, A f_l) = \delta_{kl}.
\end{equation}
(Note that it may be convenient to consider a complex eigenbasis. Hence we complexify this real Hilbert space in the obvious way).

We obtain the eigenfunctions of $\mc E^\dagger \mc R_{\rho}^\dagger$ by differentiating $G(f)$ with respect to the basis functions $f_k$. Indeed, let $\delta_k$ denote the functional derivative in the direction of $f_k$, i.e., for any functional $Z(f)$,
\begin{equation}
\delta_k Z(f) := \frac{\partial}{\partial t} Z(f + t f_k) |_{t=0},
\end{equation}
then we obtain
\begin{equation}
\mc E^\dagger \mc R_{\rho}^\dagger((\delta_{k_1} \cdots \delta_{k_n}G)(0)) = \eta_{k_1}\cdots \eta_{k_n}(\delta_{k_1} \cdots \delta_{k_n}G)(0).
\end{equation}

We have
\begin{eqnarray*}
&\bigave{ (\delta_{k_1} \cdots \delta_{k_n}G)(0), (\delta_{k_1} \cdots \delta_{k_m}G)(0)}_{\rho} \\
&\quad = \delta_{k_1} \cdots \delta_{k_n}\delta_{k_1}' \cdots \delta_{k_m}' e^{(f,A f')}|_{f=f'=0}.
\end{eqnarray*}
where the primed derivatives are relative to $f'$.

The functions $ (\delta_{k_1} \cdots \delta_{k_n}G)(0)$ form an orthogonal basis of a representation of the symmetric Fock space $\mc F$ built from the test functions, with scalar product $\ave{\cdot,\cdot}_{\rho}$. One can think of $G(0) = 1 \equiv \ket 0$ as the vacuum. The other eigenfunctions are obtain by acting on it with the creation operators $a_k^\dagger$ for the ``mode'' $f_k$. The commutation relations are
\begin{equation}
[a_k,a_l^\dagger] = \delta_{kl} \one.
\end{equation}

Explicitly,
\begin{equation}
a_{k_n}^\dagger \cdots a_{k_1}^\dagger  \ket 0 =  \delta_{k_1} \cdots \delta_{k_n}G(f)|_{f=0}.
\end{equation}

\section*{Appendix B: Perturbation theory}

This pictures allows one to find the principal observables around non-gaussian states by using perturbation theory. The trick is to express the information metric with respect to the perturbed state through its kernel $K$ expressed in that Fock space. This allows one to write also the map $\mc R^\dagger_{{\rho'}}$, for the perturbed state ${\rho'}$, also perturbatively as an operator in Fock space. The eigenvalue problem can then be formulated and computed to any degree using standard Feynman diagram techniques.

Let us defined the generating functions
\begin{equation}
K(f,g) = \ave{G(f),G(g)}_{\rho'}.
\end{equation}
Differentiating this functional yields the components in the Fock basis of an operator $K$ that is the kernel of the metric defined by $\rho'$ with respect to the unperturbed metric $\rho$:
\begin{eqnarray*}
\bra 0 a_{k_1} &\cdots a_{k_n} K a_{l_m}^\dagger \cdots a_{l_1}^\dagger \ket 0\\
 &= \delta_{k_1} \cdots \delta_{k_n}\delta_{l_1}' \cdots \delta_{l_m}' K(f,f')|_{f=f'=0}
\end{eqnarray*}
where $\delta_k'$ denotes derivation with respect to $f'$ in the direction $f_k$.
Similarly,
\begin{equation}
I(f,g) = \ave{G(f),G(g)}_{\rho} = e^{(f,A g)}
\end{equation}
is the generating function of the identity operator.

It will be convenient also to consider the Fock space defined from the metric at point $\mc E(\rho)$, with the same vacuum; but with creation operators $b_k^\dagger$ defined by
\begin{equation}
b_{k_n}^\dagger \dots b_{k_1}^\dagger \ket 0 \equiv \delta_{k_n} \dots \delta_{k_1} Q(f) |_{f = 0}
\end{equation}
where
\begin{equation}
Q(f) = \mc R^\dagger_{\rho}(G(H^{-\frac 1 2}f)).
\end{equation}
These are indeed orthogonal since
\begin{equation}
\ave{Q(f),Q(g)}_{\mc E(\rho)} = e^{(H^{-\frac 1 2} f, A H^{\frac 1 2} g)} =e^{ (f, A g)}
\end{equation}
where we used the fact that $H$ is positive in terms of the scalar product $(\cdot, A \cdot)$.

In this basis we express
\begin{equation}
L(f,g) = \ave{Q(f),Q(g)}_{\mc E({\rho'})}.
\end{equation}
We can compute from our previous results that
\begin{equation}
\begin{split}
&\ave{\mc R^\dagger_{\rho}(G(f)),\mc R^\dagger_{\rho}(G(g))}_{\mc E({\rho'})} \\
& \quad \quad = \ave{G(Hf),G(Hg)}_{\rho'} e^{h^2(Hf,X^{-2} Hg)}.
\end{split}
\end{equation}

Hence we find that $K$ and $J$ are related by
\begin{equation}
L(f,g) = K(H^{\frac 1 2} f,H^{\frac 1 2} g) \,e^{h^2(H^{\frac 1 2} f,X^{-2} H^{\frac 1 2}  g)} .
\end{equation}
Also, the channel $\mc E^\dagger$ naturally maps between the two Fock spaces, represented as the operator $E$ generated by
\begin{eqnarray*}
E(f,g) &= \ave{G(f),\mc E^\dagger(Q(g))}_{\rho} = \ave{G(f), G(H g)}_{\rho}\\
& = e^{(f, A H^{\frac 1 2} g)} = e^{(H^{\frac 1 2} f, A g)}.
\end{eqnarray*}

Finally, the unknown is the operator $R$ representing $\mc R^\dagger_{{\rho'}}$ as
\begin{equation}
R(f,g) = \ave{Q(f), \mc R_{\rho'}^\dagger(G(g))}_{\mc E(\rho)}.
\end{equation}
It is defined by the relation
\begin{equation}
\ave{\mc R_{\rho'}^\dagger(G(f)),Q(g)}_{\mc E({\rho'})} = \ave{G(f),\mc E^\dagger(Q(g))}_{{\rho'}}
\end{equation}
for all $f$ and $g$.
Expanding this relation in the respective Fock basis, we obtain
\begin{equation}
R^T L = K E
\end{equation}
or
\begin{equation}
R = L^{-1} E K.
\end{equation}

If we expand
\begin{equation}
{\rho'} = \rho( \one + \lambda X_1 + \mc O(\lambda^2) ),
\end{equation}
we have
\begin{equation}
K = \one + \lambda K_1 + \mc O(\lambda^2),
\end{equation}
and
\begin{equation}
L = \one + \lambda L_1 + \mc O(\lambda^2).
\end{equation}
Expanding $ER$, we obtain
\begin{eqnarray*}
ER &= E (I + \lambda L_1 + \dots)^{-1} E (I + \lambda K_1 + \dots) \\
&= E^2 + \lambda (E^2 K_1 - E L_1 E) + \dots.
\end{eqnarray*}
In order to compute the first order corrections to the unperturbed eigenvalue problem, we need the generating function of the perturbation
\begin{equation}
V_1 = E^2 K_1 - E L_1 E.
\end{equation}
Since $E$ is diagonal in the Fock basis, we only need to worry about $K_1$ and $L_1$ directly.

The generating function $K_1(f,f')$ of the operator $K_1$ is
\begin{equation}
K_1(f,f') = \bigave{X_1 G(f)G(f')}_{\rho}.
\end{equation}
We will consider an interaction of the form
\begin{equation}
H_I = \frac 1 {4!} \sum_x \phi(f_x)^4  = \frac 1 {4!} \sum_x \frac{\delta^4}{\delta_{f_x}^4}\,e^{\phi(f)}|_{f=0},
\end{equation}
where the functions $f_x$ possibly form a different basis than $f_k$.
This generates the state ${\rho'} = \rho(\one + \lambda X_1 + \dots)$ where
\begin{equation}
X_1 = H_I - \ave{H_I}_{\rho}\one.
\end{equation}
We have
\begin{equation}
K_1(f,f') = \frac 1 {4!} \sum_x \frac{\delta^4}{\delta_{f_x''}^4} Z(f,f',f'')|_{f''=0} - \ave{H_I}_{\rho} I(f,f')
\end{equation}
where
\begin{eqnarray*}
&Z(f,f',f'') = \bigave{e^{\phi(f'')}G(f)G(f')}_{\rho} \\
&\quad = \bigave{e^{\phi(f''+f'+f)}}_{\rho} e^{-\frac 1 2 (f,A f)-\frac 1 2 (f',A f')}\\
&\quad = e^{\frac 1 2 (f''+f'+f,A(f''+f'+f)) -\frac 1 2 (f,A f)-\frac 1 2 (f',A f')}
\end{eqnarray*}
When differentiated, this free partition function yields Feynman diagrams with no propagation between the vertices associated with $f$ or $f'$ respectively.

As an example, we performed this calculation for the state corresponding to the euclidean form of the Klein-Gordon scalar field theory with $\phi^4$ interaction (in an arbitray number of spatial dimensions). The channel defined an operator $X$ which performs a gaussian convolution over scale $\sigma$ as in the quantum example in the article. We also use $Y = y^2\one$.

The free theory yields the gaussian state defined by the operator $A$, inverse of $(A^{-1}f)(x) = \beta [ \sum_i \partial_i^2 + m^2 ] f(x)$. Given that it commutes with the operator $X$ defined by
\begin{equation}
(Xf)(x) = \frac{1}{(2\pi \sigma^2)^{\frac d 2 }}\int e^{-\frac{(x-y)^2}{2\sigma^2}} f(y)\, dy,
\end{equation}
with $d$ the number of dimensions, $H$ is codiagonal with $X$ and $H$ which are all self-adjoint in the $L^2(\mathbb R)$ scalar product. 

Using the eigenfunctions of $A$:
\begin{equation}
f_k(x) = \omega_k e^{i k x}
\end{equation}
with 
\begin{equation}
\omega_k^2 = k^2 + m^2,
\end{equation}
we obtain the eigenvalues of $H$:
\begin{equation}
\eta_k = \frac{1}{1 + y^2 \omega_k^2 e^{\sigma^2 k^2}}.
\end{equation}
The normalised unperturbed ``one-particle'' principal observables are $a_k^\dagger \ket 0 \equiv \phi(f_k)$. 
Note that we used complex eigenfunctions because it makes the calculations much simpler. The degeneracy between the $k$ and $-k$ eigenfunctions allows one to recover the real eigenfunctions by linear composition of the complex ones.

The interaction is defined as above using the improper basis
\begin{equation}
f_x(y) = \delta(x-y).
\end{equation}
Hence, besides $A_{kl} = \delta(k-l)$, we find
\begin{equation}
A_{xy} := (f_x,A f_y) = \int dk \frac{e^{i k(y-x)}}{\omega_k^2}.
\end{equation}
and
\begin{equation}
A_{kx} := (f_k,A f_y) =\frac { f_k(x)} {\omega_{k}^2}.
\end{equation}

The principal observables around $\rho'$, obtained by perturbation from the one-particle observables for the gaussian state $\rho$, are $\ket \psi_k = a_k^\dagger \ket 0 + \lambda \ket{\psi_k^1} + \mc O(\lambda^2)$ where the only non-zero components of the first order correction $\ket{\psi_k^1}$ are 
\begin{eqnarray*}
\bra 0 a_{l_1} a_{l_2} a_{l_3} \ket{\psi_k^1} &=\frac{\delta(k - l_1 - l_2 - l_3)}{\omega_k \omega_{l_1} \omega_{l_2} \omega_{l_3} } \frac{(1+\eta_k)\eta_{l_1}\eta_{l_2}\eta_{l_3}}{\eta_k - \eta_{l_1} \eta_{l_2} \eta_{l_3}}\\
&+  \frac{\delta(k - l_1) \delta(l_2 + l_3)}{ \omega_{l_2} \omega_{l_3}} \frac{ \eta_{l_2} \eta_{l_3}}{(1 - \eta_{l_2} \eta_{l_3})} \int dk' \frac{1}{ \omega_{k'}^2}\\
&+ \dots
\end{eqnarray*}
where we omitted the next two terms which are obtained by rotating $l_1$, $l_2$ and $l_3$.

\section*{Appendix C: Quantum gaussian states}

We use the notation introduced in the paper. In order to compute the metric explicitly, we need the commutation relation
\begin{equation}
e^{-i \Phi(\overline f) }e^{i \Phi(g) } = e^{\frac i 2 (f,\Delta g)} \,e^{i \Phi(g - \overline f) },
\end{equation}
where we are now working in a complexified phase-space so as to accommodate imaginary time evolutions. 
Indeed, we need the group of complex matrices $s \mapsto R_s^A$ associated with the gaussian state ${\rho}$ such that
\begin{equation}
\label{imagevol}
{\rho}^s \Phi(f) {\rho}^{-s} = \Phi(R^A_s f).
\end{equation}
The metric (in the Heisenberg picture) is $\ave{A,B}_{\rho} = \int_0^1 \ave{A {\rho}^s B {\rho}^{-s}}_{\rho}$.
This group is symplectic: $R_s^T \Delta R_s = \Delta$ and leaves the state ${\rho}$ invariant: $R_s^T A R_s = \Delta$. 

Using these properties, we find that the polynomials generated by $G_A(f) = e^{\frac 1 2 (\overline f, Af) + i \Phi(f)}$ are orthogonal with respect to the metric when they are of different degrees in the canonical observables. This follows from the fact that $\ave{G_A(f),G_A(g)}_{{\rho}} = \int_0^1 e^{(f,K_s g)} ds$ where $K_s = R_{\frac s 2}^\dagger (A+ \frac i 2 \Delta) R_{\frac s 2}$. Indeed, the derivatives of the integrand evaluates to zero at $f=g=0$ whenever the number of differentiations with respect to $f$ is not equal to the number of differentiations with respect to $g$.

Moreover, since the channel maps ${\rho}$ to a gaussian state defined by the new matrix $B = X^T A X + Y$, we find that $\mc E^\dagger(G_{B}(f)) = G_A(X f)$. Finally, using the definition of $\mc R_{\rho}^\dagger$ as adjoint of $\mc E^\dagger$, we obtain 
\begin{equation}
\ave{\mc R^\dagger_{\rho}(G_A(f)),G_B(g)}_{\mc E({\rho})} = \ave{G_A(f),G_A(X g)}_{\rho}.
\end{equation}
 This imply that a polynomial of order $n$ generated by $G_A$ is mapped by $\mc R^\dagger_{\rho}$ to a polynomial of order $n$ generated by $G_B$, which is then mapped back by $\mc E^\dagger$ to a polynomial of order $n$ generated by $G_A$. 

Therefore, we conclude that the principal observables are polynomials generated by $G_A$. Finding the exact polynomials of a given degree can be done for each order independently, which is a finite-dimensional problem.

In a situation where certain modes are decoupled in both $\rho$ and $\mc E(\rho)$, such as in the quantum Klein-Gordon field example given in the text, the same argument shows that $\mc E^\dagger \mc R_\rho^\dagger$ leaves invarient the space of operators spanned by products of field operators on some fixed set of modes $k_1,\dots,k_n$. For the Klein-Gordon field, this sector is of dimension $2^n$. We can directly compute the components of $\mc E^\dagger$ in this sector as the matrix $E$ by differentiating Equ.~(\ref{eonweyl}). Using $X$ defined in Equ.~(\ref{Xdef}), we see that $E = e^{-\frac 1 2 \sum_i k_i^2 \sigma^2} \one$. Moreover, the components $K'$ of the coarse-grained metric in this sector become independent of $\sigma$ as $\sigma \rightarrow \infty$. If we write $K$ for the components of the metric itself, which does not depend on $\sigma$, we see that the components of $\mc E^\dagger \mc R_\rho^\dagger$, which are given by the matrix $E (K')^{-1} E^T K$, are asymptotically proportional to $e^{-\sum_i k_i^2 \sigma^2}$. 
Also, writing $y_\phi y_\pi = y^2$ and $y_\phi/y_\pi = \delta$, 
we obtain
\begin{equation}
\label{Kapprox}
K' = \bigotimes_{i=1}^n \left[{ y^2
\begin{pmatrix} 
\delta^2 & 0 \\ 
0 & \delta^{-2} 
\end{pmatrix} 
+ \mc O(y^0) }\right].
\end{equation}
It follows that the components of $\mc E^\dagger \mc R_\rho^\dagger$ are of order $\mc O(y^{-2n})$.

The above analysis is also true for all quantum generalisations of the Fisher information metric.
Classically, the Fisher information metric is characterised as the only metric on the manifold of probability distributions which contracts under the action of any stochastic map. In the quantum case, Petz and Sud\'ar~\cite{petz96} characterised all contractive metrics. They are defined by an {\em operator monotone} function $\theta: \mathbb R^+ \rightarrow \mathbb R^+$ such that $\theta(t) = t \theta(t^{-1})$ for all $t>0$, and $\theta(1) = 1$. An operator monotone function has the property that, when applied to operators via functional calculus, $\theta(A) \le \theta (B)$ whenever $A < B$ (i.e., $B - A$ is positive). 
The function $\theta$ defines the kernel $\Omega^{-1}_\rho$ via its inverse $\Omega_\rho$ as follows:
\begin{equation}
\Omega_\rho = r_\rho^{\frac 1 2} \,\theta(\ell_\rho r_\rho^{-1}) r_\rho^{\frac 1 2},
\end{equation}
where $r_\rho(A) := A \rho$ and $\ell_\rho(A) := \rho A$ for any matrix $A$.

The operator $\ell_{\rho} r_\rho^{-1}$ implements imaginary time evolution. Hence, using Equ.~\ref{imagevol}, and for the state $\mc E(\rho)$, it is represented by $R_1^{B} = \one + \mc O(y^{-2})$. This implies that, to lowest order in $y^{-1}$, one can use $\Omega_\rho(A) \simeq A \rho$.
The limit $y \gg 1$ also corresponds to the high temperature limit for the Klein-Gordong field, in which the covariance matrix converges to the classical one.  
One can then check that $K'$ has the form given by Equ.~\ref{Kapprox} and the argument follows.
 

\bibliographystyle{unsrt}

\bibliography{renormalization_primer}

\begin{thebibliography}{10}

\bibitem{wilson:1974a}
K.~G. Wilson and John~B. Kogut.
\newblock The renormalization group and the epsilon expansion.
\newblock {\em Phys. Rept.}, 12:75--200, 1974.

\bibitem{wilson:1975a}
Kenneth~G. Wilson.
\newblock The renormalization group: critical phenomena and the {K}ondo
  problem.
\newblock {\em Rev. Mod. Phys.}, 47(4):773--840, 1975.

\bibitem{fisher:1998a}
Michael~E. Fisher.
\newblock Renormalization group theory: its basis and formulation in
  statistical physics.
\newblock {\em Rev. Modern Phys.}, 70(2):653--681, 1998.

\bibitem{barenblatt:1996a}
Grigory~Isaakovich Barenblatt.
\newblock {\em Scaling, self-similarity, and intermediate asymptotics:
  dimensional analysis and intermediate asymptotics}, volume~14.
\newblock Cambridge University Press, 1996.

\bibitem{preskill:2000a}
John Preskill.
\newblock Quantum information and physics: some future directions.
\newblock {\em J. Mod. Opt.}, 47(2):127--137, 2000.

\bibitem{kadanoff:1976a}
Leo~P. Kadanoff.
\newblock Notes on {M}igdal's {R}ecursion {F}ormulas.
\newblock {\em Ann. Phys.}, 100:359--394, 1976.

\bibitem{kadanoff:1966a}
Leo~P. Kadanoff.
\newblock Scaling laws for {I}sing models near ${T}_c$.
\newblock {\em Physics}, 2(6):263--272, 1966.

\bibitem{kadanoff:1977a}
Leo~P. Kadanoff.
\newblock The application of renormalization group techniques to quarks and
  strings.
\newblock {\em Rev. Mod. Phys.}, 49:267--296, 1977.

\bibitem{zamolodchikov:1986a}
A.~B. Zamolodchikov.
\newblock ``{I}rreversibility'' of the flux of the renormalization group in a
  $2${D} field theory.
\newblock {\em JETP Lett.}, 43(12):730--732, 1986.
\newblock {T}ranslated from Pis$'$ma Zh. \`Eksper. Teoret. Fiz. {\bf 43},
  565--567 (1986), no. 12, ({R}ussian).

\bibitem{apenko:2012a}
S.~M. Apenko.
\newblock Information theory and renormalization group flows.
\newblock {\em Physica A}, 391(12):62--77, 2012.

\bibitem{brody:1998a}
Dorje~C. Brody and Adam Ritz.
\newblock On the symmetry of real-space renormalization.
\newblock {\em Nucl. Phys. B}, 522(3):588--604, 1998.

\bibitem{casini:2007a}
H.~Casini and M.~Huerta.
\newblock A $c$-theorem for entanglement entropy.
\newblock {\em J. Phys. A}, 40(25):7031--7036, 2007.

\bibitem{gaite:1996a}
Jos{\'e} Gaite and Denjoe O'Connor.
\newblock Field theory entropy, the ${H}$ theorem, and the renormalization
  group.
\newblock {\em Phys. Rev. D}, 54(8):5163--5173, 1996.

\bibitem{heckman13}
Jonathan~J Heckman.
\newblock Statistical inference and string theory.
\newblock {\em arXiv:1305.3621}, 2013.

\bibitem{machta13}
Benjamin~B Machta, Ricky Chachra, Mark~K Transtrum, and James~P Sethna.
\newblock Parameter space compression underlies emergent theories and
  predictive models.
\newblock {\em Science}, 342(6158):604--607, 2013.

\bibitem{hiai:1991}
Fumio Hiai and D{\'e}nes Petz.
\newblock The proper formula for relative entropy and its asymptotics in
  quantum probability.
\newblock {\em Communications in mathematical physics}, 143(1):99--114, 1991.

\bibitem{petz:1996a}
D{\'e}nes Petz.
\newblock Monotone metrics on matrix spaces.
\newblock {\em Linear algebra and its applications}, 244:81--96, 1996.

\bibitem{ohya04}
M.~Ohya and D.~Petz.
\newblock {\em Quantum entropy and its use}.
\newblock Springer Verlag, 2004.

\bibitem{rovelli14}
Carlo Rovelli and Francesca Vidotto.
\newblock {\em Covariant Loop Quantum Gravity: An Elementary Introduction to
  Quantum Gravity and Spinfoam Theory}.
\newblock Cambridge University Press, 2014.

\bibitem{beny12}
C.~B{\'e}ny and T.J. Osborne.
\newblock {\em arXiv:1206.7004}, 2012.

\bibitem{petz96}
D{\'e}nes Petz and Csaba Sud{\'a}r.
\newblock Geometries of quantum states.
\newblock {\em Journal of Mathematical Physics}, 37(6):2662--2673, 1996.

\end{thebibliography}

\end{document}